Full length article

# Effective gain spectra vary the pulses' spectrum in a mode-locked cavity

Xun Xie [a], Wonkeun Chang [b]*

[a] School of Astronautics, Harbin Institute of Technology, Harbin, 150001, China

[b] School of Electrical and Electronic Engineering, Nanyang Technological University, Singapore 639798, Singapore

A R T I C L E   I N F O

A B S T R A C T

Filters typically play a crucial role in generating solitons by strictly confirming the working wavelength in a mode-locked laser. However, we have found that the broad smoothing filter with saturable gain also helps to regulate the pulses' operating wavelength. The effective gain, formed by the original gain spectrum and filters, shifts its center based on pulse energy, thereby affecting the pulse spectrum. A virtual mode-locked cavity is established to verify this approach in simulation. The results indicate that the spectrum changes with pulse energy, leading to a wider spectrum output. Our method paves the way for further investigations of the pulse spectrum and provides a practical approach to generating pulses at specific wavelengths.

## 1. Introduction

To create a soliton, an optical cavity must achieve a balance among various factors, including loss, dispersion, nonlinearity, and gain. Due to unavoidable loss processes such as scattering, absorption, dispersion, and so on, weak pulse energy necessitates gain to compensate for losses during pulse propagation. The development of new gain media and the exploration of their properties remain a significant focus in the field of optics.

Supported by the laser rate equation, the evolution trace of pulse amplification has been theoretically studied with differential equations in the past years ago. The character and function of the general gain medium have been further investigated, as well as laser theory. Various gain effects, such as gain saturation, gain dispersion [1], gain narrowing [2,3], gain phase modulation [4], and so on, have been theoretically analyzed and have shown good agreement with experimental findings.

The gain not only supports the soliton's operation but also determines the operational wavelength within an optical fiber cavity. The original gain spectrum is produced by various rare-earth ions, including thulium, erbium, and ytterbium. Although each ion has several emission peaks, the resulting lasers are restricted to a few wavelengths, which may not satisfy the wavelength requirements for various applications.

Fortunately, there are various technologies available to change the operational wavelength of pulsed lasers. In addition to frequency generation through nonlinear processes like self-phase modulation (SPM), four-wave mixing, Raman lasers, and Brillouin lasers, filters can be an effective approach to adjust the output wavelength for low-power lasers. Common types of filters include the Lyot filter [5,6], microring resonators [7,8], fiber grating filter [9,10], fiber tunable filters [11,12], Sagnac loop filter[13,14], and so on. These filters not only select a specific wavelength range but also help reshape the gain spectrum, such as gain flattening tech [15,16]. Therefore, filters can be seen as tools for modifying the original gain, ultimately influencing the effective gain spectrum.

In this paper, we present a method for broadening the pulse spectrum by adjusting the effective gain spectrum. Our theoretical analysis shows that pulses generally move towards the central wavelength of the gain spectrum, which can change depending on the saturation level. To simulate the spectral variation process, a virtual cavity has been established. A spectrum range exceeding 40 nm can be achieved at a wavelength of 1.8 μm. This research provides a new strategy for obtaining the desired frequency in pulse lasers and contributes to a deeper understanding of the dynamic processes of pulses.

## 2. The method

The soliton is sustained and affected by a gain function. In the frequency domain, the differential equation can be represented as

$$\frac{\partial U(\omega,z)}{\partial z} = \frac{1}{2} g_{sat} g_\omega(\omega) U(\omega,z) \quad (1)$$

$U$ represents the soliton envelope in the frequency domain $U(\omega,z)=F[u(t,z)]$, and $g_{sat}$ denotes the saturation-related gain factor, and $g_\omega$ is the normalized gain function. Additionally, the time in a frame follows the group velocity of pulses, and the frequency frame is also around the soliton center frequency. We set the initial frequency to be zero and suppose the gain center is $\omega_{gain}$ away from the initial frequency of the pulse. The gain function has a parabolic curve defined by the gain bandwidth $g_{bw}$.

$$g_\omega(\omega) = 1 - \frac{1}{g_{bw}^2}(\omega - \omega_{gain})^2 \quad (2)$$

The differential function can be written as

$$\frac{\partial U}{\partial z} = \frac{g_{sat}}{2}\left(1 - \frac{\omega_{gain}^2}{g_{bw}^2}\right)U + \frac{g_{sat}}{g_{bw}^2}\left(\omega\omega_{gain} - \frac{1}{2}\omega^2\right)U \quad (3)$$

The first term on the right side is an envelope times a constant, which does not depend on frequency. It is assumed that after a d$z$ propagation, the soliton in the frequency domain transforms into

$$U(\omega, z+dz) = \exp\left(\frac{1}{2}g_{sat}\left(1-\frac{\omega_{gain}^2}{g_{bw}^2}\right)dz\right) \cdot \exp\left(g_{sat}\frac{g_{sat}}{g_{bw}^2}\left(\omega\omega_{gain}-\frac{1}{2}\omega^2\right)dz\right)U(\omega,z) \quad (4)$$

The initial pulse at $z=0$ is assumed to be a typical conventional soliton, represented by a sech function. Its frequency envelope can be expressed as:

$$U(\omega,0) = F[\text{sech}(t)] = \pi \text{sech}\left(\frac{\pi}{2}\omega\right) \quad (5)$$

The movement of the peak point in the spectrum can be seen as a change in the frequency domain. Applying the extremum condition of $\partial U(\omega, dz)/\partial \omega = 0$, we have

$$\frac{g_{sat}}{g_{bw}^2}dz(\omega_{gain} - \omega) - \frac{\pi}{2}\tanh\left(\frac{\pi}{2}\omega\right) = 0 \quad (6)$$

To ensure the calculated accuracy, the step length d$z$ is limited. Therefore, the soliton shift occurs near its center, specifically at zero frequency. The Taylor expansion $\tanh(x) \approx x$ can be applied as an approximation. Hence, the spectral variation is around:

$$d\omega\bigg|_{\frac{\partial U}{\partial \omega}=0} \approx \frac{g_{sat}\omega_{gain}dz}{g_{sat}dz + \frac{\pi^2}{4}g_{bw}^2} \quad (7)$$

Therefore, the direction of spectral variation is decided by the $\omega_{gain}$. If $\omega_{gain}$ is a positive value, the soliton will experience a blue shift. Otherwise, if $\omega_{gain}$ is not positive, the soliton will have a red shift towards the gain center at a negative frequency.

We focus on the trace of the spectrum by solving equation (1). For instance, a dechirped sech soliton sech(t), with unit intensity and an initial center at around 1600 nm, is amplified by a gain medium with the center at around 1558 nm. $g_{sat}$ is fixed to be 1m$^{-1}$. The calculated transmission step d$z$ of the gain medium is 0.05m, $g_{bw}$ is 10 THz. In order to avoid disturbances, any other effect is ignored, such as SPM, group velocity dispersion (GVD), and so on. The trace of the pulse is shown in Fig. 1.

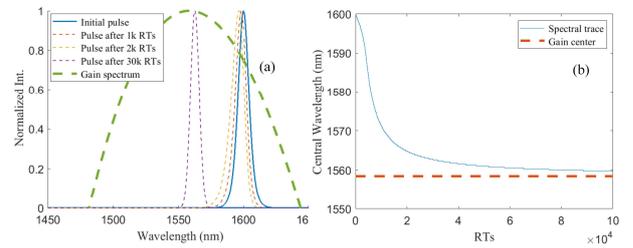

**Fig. 1. The variation of the normalized soliton in the frequency domain with an off-center gain function. (a) The location of pulses after different round-trips (RT) and the gain spectrum. (b) The trace of the pulse spectrum during RTs and the center of the gain spectrum.**

As shown, the pulse unidirectionally shifts towards the center of a gain medium, regardless of its initial wavelength. This transient process can align the pulse's wavelength with the central wavelength of the gain function. However, the shape and center of the effective gain function, after being modified by filters, could change with different parameters. This could result in a shift in the spectrum during each RT.

Different from the narrow bandwidth filter, which strictly restricts pulses' working wavelength, this filter is specifically designed to have a wide bandwidth and a smoothing filter effect. The effective gain spectrum is generated by overlaying the filter function with the original gain spectrum.

Assuming the filter follows a Gaussian function, it can be written as:

$$filter(\omega) = \exp\left(-\frac{1}{2f_{bw}^2}(\omega - \omega_{filter})^2\right) \quad (8)$$

$f_{bw}$ is the filter bandwidth. Meanwhile, the effective gain can be written as the gain function times the filter function by simply ignoring the saturation variation in the length of the gain medium.

$$G_{eff} \approx \exp\left(\frac{g_{sat}}{2}\left(1 - \frac{1}{g_{bw}^2}(\omega - \omega_{gain})^2\right)L_{gain} - \frac{1}{2f_{bw}^2}(\omega - \omega_{filter})^2\right) \quad (9)$$

Applying the condition of $\partial G_{eff}/\partial\omega=0$, the center of the effective gain $\omega_{eff}$ can be written as:

$$\omega_{eff}\bigg|_{\frac{\partial G_{eff}}{\partial \omega}=0} = \frac{g_{sat}L_{gain}(f_{bw}/g_{bw})^2 \omega_{gain} + \omega_{filter}}{g_{sat}L_{gain}(f_{bw}/g_{bw})^2 + 1} \quad (10)$$

Therefore, the center of the effective gain spectrum is a parameter related to the term of $g_{sat}L_{gain}(f_{bw}/g_{bw})^2$. If the term is very large, the center of the effective gain spectrum is close to the spectrum of the gain medium. Conversely, when the product is close to zero, the effective gain spectrum can be the same as the spectrum of the filter. The parameters $g_{bw}$, $L_{gain}$, and $f_{bw}$ are fixed in each RT, yet the $g_{sat}$ can change with the energy of pulses. Therefore, not only does the intensity of the gain spectrum change with the saturation level, but it also affects the position of the peak of the effective gain spectrum.

The pulse spectrum is proven to move towards the center of gain from equation (7). Consequently, the pulse spectrum could have a red shift or blue shift, which is decided by the center shift of the effective gain spectrum, as illustrated in Fig. 2.

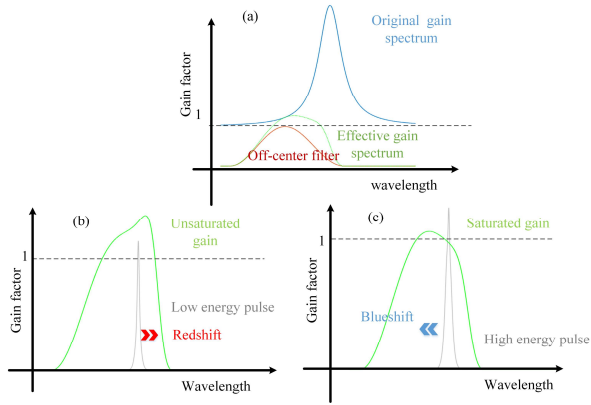

**Fig. 2. (a) The effective gain spectrum can be achieved by an original gain spectrum and an off-center smoothing broad filter. (b) The unsaturated gain spectrum drives the pulse to a red shift. (c) The saturated gain spectrum drives the pulse to a blue shift.**

In order to achieve a varying effective gain spectrum, the pulse energy should drastically fluctuate to significantly change the saturation-related factor $g_{sat}$. We propose an approach to achieve the drastic fluctuation as follows. There might exist other situations to work out a similar effect.

Energy fluctuations can be generated in an anomalous dispersion cavity, where pulse fission may occur due to excessive gain or nonlinear effects. The lower-energy components of the fission pulses are then absorbed by a mode-locking mechanism. Since the fission occurs endlessly with the accumulation of nonlinear phase shifts, the pulse energy never stabilizes. At the same time, the gain saturation could limit the maximum pulse energy, helping to maintain a relative balance within the cavity. By carefully designing the saturation level, the pulse can exhibit a fluctuating spectrum. This saturation level can be modified by adjusting the concentration of the doped rare-earth ion or the length of the fiber in experiments set up.

## 3. Simulation

To further investigate this effect, a virtual mode-locking cavity is established to produce the process, as shown in Fig. 3. This oscillator is made up of two types of fibers: a 10 m average transmission fiber with GVD of -50 ps²/km and a nonlinear index of 2.61 W⁻¹km⁻¹, and a 5 m saturable gain fiber with a GVD of 30 ps²/km and a nonlinear index of 9.08 W⁻¹km⁻¹. The gain fiber can provide saturated gain when the pulses achieve a relatively high energy level. Following the gain fiber, there is a saturable absorber, an off-center filter, and a 25% coupler. It is assumed that the filter has a Gaussian profile, while the original gain spectrum follows a Lorentzian shape. In addition to the parameters mentioned above, the other parameters of the intracavity components are detailed in Table 1.

Table 1. Parameters for the simulation

| Parameters | Symbols | Value |
|---|---|---|
| Gain center of the fiber | $\omega_{gain}$ | 1850 nm |
| Gain bandwidth of the fiber | $g_{bw}$ | 8 THz |
| Saturation level of the fiber | $E_{sat}$ | 300 pJ |
| Small signal gain | $g_0$ | 1.2 m⁻¹ |
| Center of the Gaussian filter | $\omega_{filter}$ | 1750 nm |
| Bandwidth of the filter | $f_{bw}$ | 10 THz |
| Saturation level of the absorber | $P_{sat}$ | 15 W |
| Modulation depth | $T_1$ | 1 |
| Loss of fibers | $\alpha$ | 1 km⁻¹ |

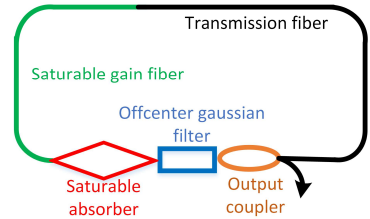

**Fig. 3. The simulated model cavity with a varying effective gain spectrum**

Here, a CGLE is used to simulate the pulse propagation. The equation can be written as follows, and a split-step Fourier method is applied to carry out the simulation. All parameters remain consistent with those previously stated.

$$\frac{\partial u(t,z)}{\partial z} = -\alpha - i\beta_2 \frac{\partial^2 u}{\partial t^2} + i\gamma|u^2|u + g(\omega)u \quad (11)$$

The gain function is assumed to follow a Lorentzian curve because the spectrum of the gain medium used to be a Lorentzian curve in measurements [17]. The parabolic function is not suitable because the gain factor should always be a positive value, despite its simple computation.

$$g(\omega) = \frac{g_o}{2\left(1 + \int |u^2|/E_{sat}\,dt\right)} \frac{g_{bw}/2}{\pi\left((\omega-\omega_{gain})^2 + (g_{bw}/2)^2\right)} \quad (12)$$

The transmission function for the saturable absorber can be expressed as

$$T_{SA} = \sqrt{(1-T_1) + T_1\left(\frac{1}{1+P_{SA}/|u|^2}\right)} \quad (13)$$

The function of the Gaussian filter is the same as equation (8). To exclude the influence of the initial condition, the results of the first 2,000 roundtrips are discarded. A simulation of 10,000 RTs is conducted, with the results presented in the frequency domain and time domain in Fig. 4 and 5, respectively.

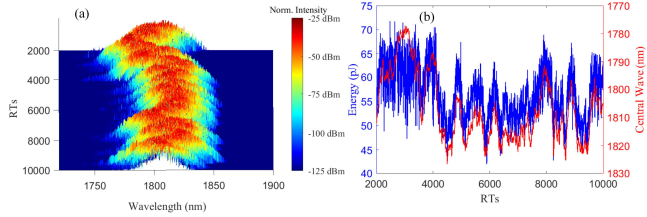

**Fig. 4. (a) The evolution trace of pulses in the frequency domain (b) Pluses' central wavelength and their total energy vary with roundtrips**

The spectra shown in Fig. 4 (a) feature several spines, making it difficult to determine their movement by simply tracing the shift of the maximum point of the spectrum. Instead, the central wavelength can be determined by calculating its mass center of each output. Fig. 4 (b) illustrates a distinct relationship between pulse energy and central wavelength. It is evident that the spectral variation spans more than 40 nm, and the pulse energy changes oppositely with the central wavelength. A correlation formula is used to quantify this relationship, which can be expressed as

$$\text{CI} = \sum_{i=1}^{n}(x_i - \overline{x})(y_i - \overline{y}) \Big/ \sqrt{\sum_{i=1}^{n}(x_i - \overline{x})^2} \sqrt{\sum_{i=1}^{n}(y_i - \overline{y})^2} \quad (14)$$

The correlation index (CI) between sequences x and y is calculated to be -0.79, indicating a strong negative correlation between the central wavelength and pulse energy. Achieving a CI of -1 is unlikely since the filter and gain function do not simultaneously affect the pulse. This index supports our method of varying pulse spectrum through the change in effective gain spectrum.

The evolution trace of the soliton profile is shown in Fig. 5. Since the central wavelength of the pulse spectrum varies with RTs, it leads to unavoidable group delays. After each propagation distance (d$z$), the pulse in the frequency domain would append a term 'exp(i$\beta_2\omega_{gain}$d$z\omega$)', which transforms into a Dirac function when the spectra are converted back to the time domain. This Dirac function convolves with the pulse function, resulting in group delays for each propagation.

To avoid the delayed pulse reaching the margin of the simulated frame, we keep the centralization of the pulse bunch by making sure the highest peak is consistently at zero time. The evolution trace of pulses in the time domain is shown in Fig. 5 (a). Typically, a piece of non-periodic trace is analyzed for 6428$^{th}$ to 6448$^{th}$ RTs in Fig. 5 (b).

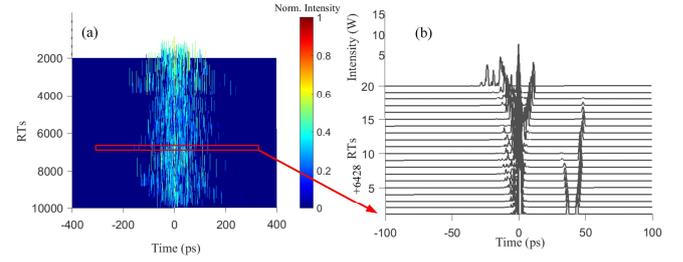

**Fig. 5. (a) The evolution trace of pulses in the time domain after centralizing the maximum peak. (b) Typical nonperiodic trace of pulses profile.**

The segment of the trace shows the pulse evolution trace in detail. As previously expected, the pulses continue to fission and dissipate in their propagation. This characteristic suggests that the pulses may operate in a soliton rain regime. The low-energy pulses gradually diminish and vanish during transmission, as observed in the pulse trace around 50 ps. In contrast, the high-energy pulse can split due to nonlinear effects. Due to the limit of the saturation level, the pulse competes with the gain. The peak power of pulses is clamped in the cavity. The pulses, whether absorbed or amplified during transmission, experience significant fluctuations that alter the effective gain through equation (10), ultimately resulting in a broad spectrum.

## 4. Conclusion

In conclusion, we conduct a theoretical exploration of a technique for broadening the pulse spectrum by creating a variable spectrum in RTs. The principle is illustrated and discussed with differential equations. A simulated oscillator is established based on a design of an effective gain spectrum, which could be constructed by a saturable gain medium and a smoothing broad bandwidth filter. The pulse behavior aligns with our expectations. We believe the technology can contribute to achieving the needed wavelength in numerous applications, and it can offer a novel method for manipulating solitons.